\documentclass[12pt,aps,prd,preprint,tightenlines,superscriptaddress,
   showpacs,nofootinbib]{revtex4}
\newcommand{\PRE}[1]{{#1}} 

\usepackage{bm}
\usepackage{graphicx}

\graphicspath{{figs/}}

\newcommand{\ipb}{\text{pb}^{-1}}
\newcommand{\ifb}{\text{fb}^{-1}}

\newcommand{\mev}{\text{MeV}}
\newcommand{\gev}{\text{GeV}}
\newcommand{\tev}{\text{TeV}}

\newcommand{\eqref}[1]{Eq.~(\ref{#1})}

\newcommand{\secref}[1]{Sec.~\ref{sec:#1}}

\newcommand{\figref}[1]{Fig.~\ref{fig:#1}}
\newcommand{\figsref}[2]{Figs.~\ref{fig:#1} and \ref{fig:#2}}
\newcommand{\Figref}[1]{Figure~\ref{fig:#1}}

\newcommand{\Dsle}[1]{\hskip 0.09 cm \slash\hskip -0.28 cm #1}
\newcommand{\met}{{\Dsle E_T}}
\newcommand{\mht}{{\Dsle H_T}}

\newcommand{\bigDsle}[1]{\hskip 0.05 cm \slash\hskip -0.38 cm #1}
\newcommand{\bigmet}{{\bigDsle E_T}}

\newcommand{\be}{\begin{equation}}
\newcommand{\ee}{\end{equation}}
\newcommand{\bea}{\begin{eqnarray}}
\newcommand{\eea}{\end{eqnarray}}
\newcommand{\gsim}{\lower.7ex\hbox{$\;\stackrel{\textstyle>}{\sim}\;$}}
\newcommand{\lsim}{\lower.7ex\hbox{$\;\stackrel{\textstyle<}{\sim}\;$}}

\begin{document}

\preprint{FERMILAB-PUB-11-319-T}
\preprint{UCI-TR-2011-10}
\preprint{UH-511-1171-11}

\title{ \PRE{\vspace*{0.8in}}
$\bm{B'}$s with Direct Decays: Tevatron and LHC Discovery Prospects in
the $\bm{b\bar{b}\bigmet}$ Channel
\PRE{\vspace*{0.3in}}
}

\author{Johan Alwall}
\affiliation{Theoretical Physics Department, Fermi National
  Accelerator Laboratory, P.O.~Box 500, Batavia, IL 60510, USA
\PRE{\vspace*{.1in}}
}

\author{Jonathan L.~Feng}
\affiliation{Department of Physics and Astronomy, University of
California, Irvine, CA 92697, USA
\PRE{\vspace*{.1in}}
}

\author{Jason Kumar}
\affiliation{Department of Physics and Astronomy, University of
Hawai'i, Honolulu, HI 96822, USA
\PRE{\vspace*{.1in}}
}

\author{Shufang Su%
\PRE{\vspace*{.4in}}
}
\affiliation{Department of Physics and Astronomy, University of
California, Irvine, CA 92697, USA
\PRE{\vspace*{.1in}}
}
\affiliation{Department of Physics, University of Arizona, Tucson, AZ
85721, USA
\PRE{\vspace*{.5in}}
}

\date{July 2011}

\begin{abstract}
\PRE{\vspace*{.3in}} We explore the discovery prospects for $B'
\bar{B}'$ pair production followed by direct decays $B' \to b X$,
where $B'$ is a new quark and $X$ is a long-lived neutral particle.
We develop optimized cuts in the $(m_{B'}, m_X)$ plane and show that
the 7 TeV LHC with an integrated luminosity of $1\ (10)~\ifb$ may
exclude masses up to $m_{B'} \sim 620 \ (800)~\gev$, completely
covering the mass range allowed for new quarks that get mass from
electroweak symmetry breaking.  This analysis is applicable to other
models with $b \bar{b} \met$ signals, including supersymmetric models
with bottom squarks decaying directly to neutralinos, and models with
exotic quarks decaying directly to GeV-scale dark matter.  To
accommodate these and other interpretations, we also present
model-independent results for the $b \bar{b} \met$ cross section
required for exclusion and discovery.
\end{abstract}

\maketitle

\section{Introduction}
\label{sec:intro}

This is an exciting time for TeV-scale colliders, with experiments at
the Tevatron and Large Hadron Collider (LHC) collecting data at
unprecedented luminosities and energies.  In this study, we explore
the prospects for discovering new physics through $B' \bar{B}'$
production, followed by the direct decays $B' \to b X$, where $B'$ is
a new down-type quark (with electric charge $q_{B'} = -{1\over 3}$)
and $X$ is a long-lived neutral particle, leading to the signal $b
\bar{b} \met$.  This study complements our previous study of $T'
\bar{T}'$ production (where $T'$ is an up-type quark with $q_{T'}=
{2\over 3}$), followed by the direct decay $T' \to t X$, leading to
the signal $t \bar{t} \met$~\cite{Alwall:2010jc}.

The possibility of new physics leading to heavy flavor signals is, of
course, well-appreciated, but such signals are usually accompanied by
other visible particles from multi-step cascade decays.  The direct
decays considered here are much less studied, but are well-motivated
from many perspectives. The gauge hierarchy problem, for example,
motivates top and bottom partners, new particles that cancel the
radiative contributions from bottom and top quark loops to the Higgs
boson mass.  The fine-tuning in such models is generally reduced when
these partners are light, making it natural that such particles are
among the lightest new particles and decay without cascades.  The
canonical example is supersymmetric models with top and bottom squarks
that are lighter than the other squarks and decay directly through
$\tilde{b} \to b \chi_1^0$ and $\tilde{t} \to t \chi_1^0$, where
$\chi_1^0$ is the lightest neutralino.

Dark matter provides another general motivation for the signals we
consider.  In many models of dark matter, the dark matter particle $X$
is the lightest particle charged under an exact symmetry, ``dark
charge,'' and it may scatter off normal matter through processes $Xq
\to Q' \to Xq$, where $Q'$ is another new particle.  This possibility
is especially motivated at present by the possibility that such
signals may in fact have been seen at DAMA~\cite{DAMA},
CoGeNT~\cite{CoGeNT}, and CRESST~\cite{CRESST}.  In such scenarios,
the $Q'$ particles are necessarily colored and have dark charge; they
can be produced through $q \bar{q} / gg \to Q' \bar{Q}'$ and decay
directly through $Q' \to q X$.  Although these decays may be to any
quark flavor, decays to $b$ and $t$ are realized in concrete scenarios
with WIMP dark matter, WIMPless dark matter, and asymmetric dark
matter, as we review in \secref{models}.  Such models are also, of
course, much more amenable to study at hadron colliders than those in
which the decays are solely to light quarks.

In a previous study~\cite{Alwall:2010jc}, we investigated the collider
reach for up-type quark pair production $q \bar{q} / gg \to
T'\bar{T}'$, followed by $T' \to t X$.  In this work, we analyze the
pair production of down-type quark $B' \bar{B}'$ with subsequent decay
of $B' \to b X$ at both the Tevatron and 7 TeV LHC using the
Madgraph/MadEvent/Pythia/PGS4 packages.  For relatively small values
of $m_X$, we will find that $B'$ masses up to 440, 460 and 480 GeV may
be excluded given integrated luminosities of 5, 10, and $20~\ifb$ at
the Tevatron, respectively.  This reach is greatly enhanced at the 7
TeV LHC: with an integrated luminosity of only $100~\ipb$, the 7 TeV
LHC's 95\% CL exclusion reach is comparable to that of the Tevatron
with $20~\ifb$.  The whole region of $m_{B'}$ allowed by Yukawa
coupling perturbativity can be explored with $1~\ifb$ of data, and,
with $10~\ifb$ of data, $B'$ masses up to 800 GeV may be excluded.
The 7 TeV LHC also has great potential in terms of $B'$ discovery:
$3\sigma$ discovery contours reach $B'$ masses of 540 and 700 GeV for
integrated luminosities of 1 and $10~\ifb$, respectively.

We also present model-independent results for collider reaches as a
function of the $bX \bar{b}X$ production cross-section, with $m_X = 1$
GeV.  From these, for any theoretical prediction for $\sigma(B'
\bar{B}')\times B (B' \to b X)^2$ as a function of $m_{B'}$, one can
easily determine the expected exclusion and discovery reaches in
$m_{B'}$.  Our results may therefore be applied to other models that
give rise to the $b\bar{b}\met$ signal.

In \secref{models}, we discuss models that yield the $b \bar{b} \met$
signal and their implications for the $B'$ and $X$ masses.  In
\secref{limits} we discuss existing bounds on these scenarios.  Our
simulation is described in \secref{collider}. The results are
presented in \secref{results} and summarized in \secref{conclusions}.
In the Appendix, we list cross sections at the Tevatron and 7 TeV LHC
for standard model (SM) backgrounds and several benchmark points after
various levels of cuts.

\section{Models}
\label{sec:models}

We now discuss models that yield the $b \bar{b} \met$ signal and their
implications for the $B'$ and $X$ masses.  We begin with the familiar
examples of supersymmetry and universal extra dimensions (UED), where
the spins of the $B'$ and $X$ particles are $(S_{B'}, S_X) = (0,
\frac{1}{2})$ and $(\frac{1}{2}, 1)$, respectively.  These are model
frameworks in which existing searches have been carried out and the
$X$ particle is WIMP dark matter.  We then discuss models with
WIMPless and asymmetric dark matter, where the spins are $(S_{B'},
S_X) = (\frac{1}{2}, 0)$.  In these models, $X$ is again dark matter,
but the light mass range $m_X \sim 1-10~\gev$ is particularly
motivated by currently claimed signals. Strictly speaking, our
analysis is valid only for the spin assignment of the WIMPless and
asymmetric dark matter cases, but as described below, it is also
applicable to the other scenarios with minor modifications.

\subsection{Supersymmetry with Light Bottom Squarks}
\label{sec:susy}

Supersymmetric models yield the $b \bar{b} \met$ signal when bottom
squark pair production is followed by direct decays $\tilde{b} \to b
\chi_1^0$, where $\chi_1^0$ is the lightest neutralino, an excellent
dark matter candidate~\cite{Goldberg:1983nd,Ellis:1983ew}.  Squarks
are often assumed to decay through cascade decay chains.  In contrast
to other squarks, however, bottom (and top) squarks have their masses
reduced by the impact of large Yukawa couplings on renormalization
group evolution, and their masses are also directly constrained by
naturalness. There are therefore reasons to expect the bottom squarks
to be relatively light and decay directly to the lightest
supersymmetric particle, even if other squarks are heavy and decay
through cascades.

The salient features for this analysis are
\begin{itemize}
\setlength{\itemsep}{1pt}\setlength{\parskip}{0pt}\setlength{\parsep}{0pt}
\item The signal arises from bottom squark pair production followed by
 $\tilde{b} \to b \chi_1^0$.  The decaying particle is a scalar, in
 contrast to all other examples discussed below.
\item The $\tilde{b}$ mass is only constrained by direct searches
discussed in \secref{limits}, which require $m_{\tilde{b}} \agt
230~\gev$ for small $m_{\chi_1^0}$.  The mass limit on $m_{\tilde{b}}
$ becomes much weaker for small mass splitting $m_{\tilde{b}} -
m_{\chi_1^0}$.
\item The neutralino mass satisfies $m_{\chi_1^0} \agt 47~\gev$,
assuming gaugino mass unification~\cite{Nakamura:2010zzi}.  Without
gaugino mass unification, there is no lower bound on the neutralino
mass, which in general may be anywhere in the range $0 \alt
m_{\chi_1^0} < m_{\tilde{b}}$.
\end{itemize}

\subsection{Universal Extra Dimensions}

UED models give the desired signature, where the new down-type quark
$B'$ is identified with the Kaluza-Klein bottom quark $b^1$.  The
$b^1$ can be pair-produced and then decay directly to Kaluza-Klein
hypercharge gauge bosons $B^1$, which may be WIMP dark
matter~\cite{Servant:2002aq,Cheng:2002ej}.

The salient features of this model are
\begin{itemize}
\setlength{\itemsep}{1pt}\setlength{\parskip}{0pt}\setlength{\parsep}{0pt}
\item The signal arises from $b^1$ pair production followed by $b^1
  \to b B^1$.
\item The $b^1$ mass is set by the size of the extra dimension.  Its
  mass is constrained only by the direct searches discussed in
  \secref{limits}, which require $m_{b^1} \agt 440~\gev$ for small
  $m_{B^1}$.  If $m_{B^1} \approx m_{b^1}$, the mass limit is much
  weaker.

\item The size of the extra dimension typically sets the size of all
  the Kaluza-Klein particles, and so UED spectra are typically
  expected to be compressed relative to supersymmetry. One therefore
  expects $m_{B^1} \sim m_{b^1}$, but $B^1$ masses anywhere in the
  range $0 \alt m_{B^1} < m_{b^1}$ are experimentally viable.
\end{itemize}

\subsection{WIMPless Dark Matter}

In WIMPless scenarios~\cite{WIMPless_setup}, dark matter is in a
hidden sector. These scenarios have the notable feature that the dark
matter candidate automatically has approximately the correct relic
density, regardless of the candidate particle's mass.

The WIMPless dark matter particle $X$ couples SM quarks to
new quarks through Yukawa interactions
\begin{equation}
V = \lambda^q_i X \bar Q'_L q_{L\, i} +
\lambda^u_i X \bar T'_R u_{R\, i} +
\lambda^d_i X \bar B'_R d_{R\, i} \ ,
\end{equation}
where $X$ is assumed here to be a complex scalar charged under a
discrete symmetry, $q_{L\, i}^T = (u_{L\, i} , d_{L \, i} )$, $u_{R\,
i}$, and $d_{R\, i}$ are the SM quarks of generation $i$, and
$Q_L^{\prime\, T} =(T'_L ,B'_L)$, $T'_R$, and $B'_R$ are the new
quarks, also charged under the same discrete symmetry as $X$.

In general, the Yukawa couplings can couple $X$ to any of the SM
generations, subject to flavor constraints~\cite{IVDM}.  Although it
is difficult to know what a ``natural'' flavor structure for new quark
couplings should be, one reasonable possibility is that new quark
couplings follow the observed Yukawa hierarchy and couple new quarks
dominantly to third generation quarks.  In fact, for ${\cal O}(1)$
Yukawa couplings, WIMPless models with dark matter coupled to 3rd
generation quarks may explain the reported dark matter signals from
DAMA and CoGeNT~\cite{WIMPless3rdgen}.

For the purpose of the analysis presented here, the salient features
are
\begin{itemize}
\setlength{\itemsep}{1pt}\setlength{\parskip}{0pt}\setlength{\parsep}{0pt}
\item The signal arises from $B'$ pair production followed by the
 decay $B' \to b X$. Dark charge conservation forbids the cascade
 decays $B' \to Wq$ and $B' \to Zq$, and the possibilities $B' \to dX,
 sX$ are excluded by hand.
\item $T'$ and $B'$ are new quarks that get mass through electroweak
 symmetry breaking, with $m_{T', B'} = \lambda_{T', B'} v / \sqrt{2}$,
 where $v \simeq 246~\gev$.  Yukawa coupling perturbativity requires
 $\lambda_{B'}^2 \alt 4\pi$, which implies the upper bound $m_{B'}
 \alt 600~\gev$.  For small $m_X$, direct
 searches~\cite{D0sb,CDFsb,CDFgluino} place a lower bound, $m_{B'}
 \agt 440~\gev$.  This lower bound is weakened for larger dark matter
 mass.
 \item For $X$ to freeze out with the correct relic density and
 preserve a key motivation for WIMPless scenarios, it cannot be
 extremely light, but the range $10~\mev \alt m_X < m_{B'}$ is
 allowed.  However, if one hopes to explain the DAMA~\cite{DAMA} and
 CoGeNT~\cite{CoGeNT} anomalies and be marginally consistent with
 stringent exclusion bounds from CDMS~\cite{CDMS} and
 XENON10/100~\cite{XENON}, light masses with $m_X \sim 7~\gev$ are
 preferred.
\end{itemize}

There are several other models that share the basic features described
above.  One well-known example is little Higgs models, where the new
quarks are not 4th generation quarks, but instead arise from the extra
degrees of freedom needed when the gauge symmetry ${\rm SU}(2)_L$ is
enlarged.  Unlike the WIMPless case, the mass of the new quarks is not
generated by Yukawa couplings to the SM Higgs, and thus is not bounded
from above by perturbativity.  These quarks may also be charged under
$T$-parity, decaying to SM quarks plus dark matter (the lightest
particle charged under $T$-parity)~\cite{littleHiggsT}.  Another
example is provided by a recent set of models in which ${\rm U}(1)_B$
is a gauge symmetry~\cite{U(1)B_DM}.  In this case, the new quarks are
4th generation (possibly mirror) quarks that are added to cancel the
${\rm U}(1)_B$ mixed anomaly.  They are charged under a new ${\rm
U}(1)$ global symmetry, and the dark matter is a scalar which is the
lightest particle charged under this global ${\rm U}(1)$.

\subsection{Asymmetric Dark Matter}

Another class of models predicting the $b \bar{b} \met$ signature are
models of asymmetric dark matter arising from hidden sector
baryogenesis~\cite{Dutta:2010va}.  In this framework, sphalerons of a
hidden sector gauge group generate a baryon asymmetry by producing
exotic quarks~\cite{Dutta:2006pt}.  The $B'$ is the down-type exotic
quark, which decays to dark matter and right-handed bottom quarks, $B'
\to b_R X$.  Our analysis is directly applicable when the dark matter
particle is a scalar.\footnote{In general, the dark matter candidate
can also be spin-$\frac{1}{2}$ for this model.  In this case, the
relevant signal is pair production of the down-type exotic squark,
$\tilde B'$, followed by the decay $\tilde B' \to b_R X$.}  In this
model, the number density of the dark matter candidate is determined
by the baryon number density.  Moreover, the dark matter multiplet and
$b$-quark multiplet are both chiral under a ${\rm U}(1)_{T3R}$ gauge
group; both $m_X$ and $m_b$ are determined by the symmetry-breaking
scale of ${\rm U}(1)_{T3R}$, so one expects $m_X \sim m_b$.  As a
result, this asymmetric dark matter candidate naturally has
approximately the correct relic density.

For the purposes of this analysis, the salient features of the
asymmetric dark matter model are
\begin{itemize}
\setlength{\itemsep}{1pt}\setlength{\parskip}{0pt}\setlength{\parsep}{0pt}
\item The exotic down-type quark $B'$ is not a 4th generation quark,
 and its mass is not generated by electroweak symmetry breaking.  As a
 result, there is no upper bound on its mass, which is only
 constrained by direct searches to satisfy $m_{B'} \agt 440~\gev$ for
 small $m_X$.
\item The asymmetric dark matter should have $m_X \sim 1-10~\gev$ to
correctly explain the relic density.  But a dark matter mass $m_X \sim
7~\gev$ is preferred to explain reported direct detection signals.
\item The exotic quark need not be down-type, but if it is, it
 necessarily decays through $B' \to bX$.
\end{itemize}

\section{Current Collider Limits}
\label{sec:limits}

In this section, we summarize existing constraints on the $b \bar{b}
\met$ signature and related channels. All bounds quoted below are 95\%
CL constraints.

As noted in \secref{intro}, we are interested in $B' \bar{B}'$
production followed by direct decays $B' \to b X$.  These differ from
searches for conventional fourth generation quarks, which we denote by
the lowercase $b'$ and $t'$, which typically decay through
cascades. Nevertheless, we begin with these as a useful reference
point. Searches for $t'$ and $b'$ has been performed at both Run II of
the Tevatron~\cite{Aaltonen:2009nr,Aaltonen:2011vr,CDF4tht,D04tht} and
the LHC~\cite{Chatrchyan:2011em,ATLAS4th}.  These searches assume that
the fourth generation quarks couple to the first three generations and
decay through $b', t' \to q W$. For $b'$, the most stringent result at
present is from CDF searches for $b' \bar{b}'$ production followed by
$b' \to t W$.  The lack of an excess in $4.8~\ifb$ of data implies
$m_{b'} > 372~\gev$~\cite{Aaltonen:2011vr}.  For $t'$, CDF finds no
signal for $t' \bar{t}'$ production followed by $t' \to b W$ in
$5.6~\ifb$ of data, implying $m_{t'} > 358~\gev$~\cite{CDF4tht}, and a
D\O\ search for $t' \bar{t}'$ followed by $t' \to q W$ in $4.3~\ifb$
of data requires $m_{t'} > 296~\gev$~\cite{D04tht}.  At the LHC, null
results from CMS searches for $b' \to t W$ using $34~\ipb$ of data
imply $m_{b'} > 361~\gev$~\cite{Chatrchyan:2011em}.  ATLAS analyses of
$t'\ \text{or}\ b' \to qW$ in $37~\ipb$ of data imply $m_{t', b'} >
270~\gev$~\cite{ATLAS4th}.  We stress again, however, that the limits
of this paragraph do not apply to the $B'$ and $T'$ searches we
consider here, as decays $B' \to q W$ and $T' \to q W$ are excluded by
dark charge conservation.

As noted in \secref{susy}, however, the $b \bar{b} \met$ signal is
produced in the case of supersymmetry with bottom squark pair
production followed by $\tilde{b}\to b \chi_1^0$, where $\chi_1^0$ is
the lightest neutralino.  Both CDF and D\O\ have searched for this
signal.  D\O\ finds no excess in $5.2~\ifb$ of data, requiring
$m_{\tilde{b}} > 247~\gev$ for $m_{\chi_1^0} =0$ and excluding
$160~\gev < m_{\tilde{b}} < 200~\gev$ for $m_{\chi_1^0}
=110~\gev$~\cite{D0sb}.  The corresponding CDF result for
$m_{\chi_1^0} = 0$ using $2.65~\ifb$ of data is $m_{\tilde{b}} >
230~\gev$~\cite{CDFsb}. Taking into account only the difference in $B'
\bar{B'}$ and $\tilde{b} \tilde{b}^*$ cross sections, the D\O\ bound
$m_{\tilde{b}} > 247~\gev$ implies $m_{B'} \agt 365~\gev$.  As we will
see in \secref{results}, an optimized collider analysis would imply
$m_{B'} \agt 440~\gev$ when cuts similar to those of the D\O\ sbottom
searches are applied to $B' \bar{B}'$ pair production.  Signal
significance would be reduced by a trials factor associated with the
choice of optimum cuts, however.

A search for gluino pair production with $\tilde{g} \to \bar{b}
\tilde{b}$ followed by $\tilde{b}\to b \chi_1^0$ has been carried out at
CDF using an integrated luminosity of $2.5~\ifb$~\cite{CDFgluino}.
Candidate events were selected requiring two or more jets, large
$\met$, and at least two $b$-tags.  Using neural net analyses, CDF
found $m_{\tilde{g}} > 350~\gev$ for large mass splittings $\Delta m
\equiv m_{\tilde{g}}-m_{\tilde{b}} \agt 80~\gev$, and about 340 GeV
for small $\Delta m \sim 20~\gev$.  The result for the case of small
$\Delta m$, where two $b$-jets are soft and sometimes missed, can be
applied to the $B'\bar{B}'$ search and imply roughly $m_{B'} \agt
370~\gev$.

SUSY searches by the ATLAS Collaboration at the LHC with
$\sqrt{s}=7~\tev$ and $35~\ipb$ luminosity~\cite{Aad:2011ks} studied
the process of gluino and sbottom pair production with $\tilde{g} \to
\bar{b} \tilde{b}_1$ and $\tilde{b}_1 \to b \chi_1^0$.  Events are
selected by requiring large $\met$ and at least three jets, of which
at least one is $b$-tagged.  For $m_{\tilde{b}_1} < 500~\gev$, this
search implies $m_{\tilde{g}} > 590~\gev$.  This limit also bounds
$B'\bar{B}' \to b\bar{b} j \met$, where the additional jet results
from QCD radiation.  It is, however, not straightforward to obtain the
mass limit on $m_{B'}$ without detailed collider analyses.

Although we focus here on $B'$ production, if the $B'$ mass is
generated by electroweak symmetry breaking then the mass difference
between $B'$ and $T'$ is constrained by electroweak precision data to
be less than about 50 GeV~\cite{4thgeneration}.  Therefore, bounds on
$T'$ production are also relevant.  The discovery prospects for $T'
\bar{T}'$ production followed by direct decays were evaluated in
Ref.~\cite{Alwall:2010jc}, and we summarize current bounds here.

A CDF search for $T'\bar{T'} \to t\bar{t} \met$ in the semi-leptonic
channel in $4.8~\ifb$ of data implies $m_{T'} > 360~\gev$ for $m_X
\leq 100~\gev$~\cite{Aaltonen:2011rr}.  SUSY searches for stop pair
production at the Tevatron followed by $\tilde{t}\to b \ell \tilde\nu$
also imply bounds on $T'\bar{T'} \to t\bar{t} \met$ when both tops
decay leptonically.  Null results from searches at both
CDF~\cite{Aaltonen:2010uf} and D\O~\cite{Abazov:2008kz} in $1~\ifb$ of
data imply $m_{\tilde{t}} \agt 180~\gev$ for $m_{\tilde\nu} \alt
100~\gev$.  Accounting for the difference in $\tilde{t} \tilde{t}^*$
and $T' \bar{T}'$ cross sections, this implies the bound $m_{T'} \agt
263~\gev$.  The CDF Collaboration has also reported a search for top
squark pair production based on an integrated luminosity of
$2.7~\ifb$, using the purely leptonic final states from $p\bar{p} \to
\tilde{t}_1\tilde{t}_1^*$, followed by $\tilde{t}_1\to b \chi_1^\pm
\to b \ell \nu \chi_1^0$~\cite{CDFstop}.  The data are consistent with
the SM background, leading to the constraint $m_{\tilde{t}_1} \agt 150
- 185~\gev$, where the exact limit depends on $m_{\chi_1^0}$,
$m_{\chi_1^\pm}$, and $B (\chi_1^\pm\to l^\pm \nu \chi_1^0)$.

Stop pair production (either direct or via gluino decay $\tilde{g} \to
\bar{t} \tilde{t}_1$) has been searched for at ATLAS with
$\sqrt{s}=7~\tev$ and $35~\ipb$~\cite{Aad:2011ks}.  Assuming $B (
\tilde{t}_1 \to b \chi_1^\pm) = 1$ and $B(\chi_1^\pm \to
\chi_1^0 W^{(*)}) = 1$, searches have been performed in the
semi-leptonic channel with 1 lepton, 2 jets (including one $b$-jet)
and large $\met$.  For $130~\gev < m_{\tilde{t}} < 300~\gev$, this
search implies $m_{\tilde{g}} > 520~\gev$. Cross sections for
$\tilde{g}\tilde{g} + \tilde{t}_1 \tilde{t}_1^*$ around 8 to 40 pb
have been excluded for $400~\gev < m_{\tilde{g}} < 600~\gev$.  It is
less straightforward to translate this search limit to the $T'$ case,
given the very different cut efficiencies of the dominant
$\tilde{g}\tilde{g}$ process.

Finally, a recent search at ATLAS with $\sqrt{s}=7~\tev$ and
$35~\ipb$~\cite{ATLASTpri} for pair production of fermionic top
partners decaying to a top quark and a long-lived neutral particle
gives a mass limit of $m_{T'}>275\ (300)~\gev$ for $m_X < 50\
(10)~\gev$.  This limit can be directly applied to the case of $T'$
pair production followed by direct decays.

\section{Collider Analysis}
\label{sec:collider}

\subsection{Signal and background simulation}

Both signal and backgrounds were simulated using
MadGraph/MadEvent~4~\cite{MadGraph} and passed through Pythia
6.4~\cite{PYTHIA} (with $p_T$-ordered showers) for parton showering
and hadronization. We used the CTEQ6L1 parton distribution
functions~\cite{Pumplin:2002vw} and the factorization and
renormalization scales were set to $m^2+p_T^2$ of the massive
particles produced. Detectors were simulated with PGS4~\cite{PGS}
using the Tevatron and ATLAS detector cards for Tevatron and LHC
simulations, respectively, as provided by MadGraph/MadEvent.

Our signal process is $p p / p \bar{p} \to B' \bar{B}' + \text{jets}$
with $B'\to b X$, using matrix elements with jet matching for up to 2
jets and the decay at matrix element level, giving a signal of two
$b$-jets and missing energy, plus possible associated jets from QCD
radiation. We generated events at grid points in the $(m_{B'}, m_X)$
plane with 25,000 events per grid point.

The main backgrounds to this process are $W^\pm+$ jets, $Z+$ jets, and
$t\bar t$ production. The former two were simulated with up to 3 jets
coming from the matrix element and the latter with up to 1 jet from
the matrix element, to ensure that the backgrounds were properly
modeled with respect to the jet cuts used in the analyses. Also the
single top background and diboson background were simulated and found
to be negligible, as expected. The backgrounds have been compared to
similar Tevatron~\cite{Abazov:2010wq} and
ATLAS~\cite{ATLAS7,ATLAS10,Aad:2009wy} analyses, with agreement at the
expected 20\% level. The exception is $b$-tagging, where the
experimental efficiency for the Tevatron D\O~detector is better than
that given by PGS. We have therefore applied a correction factor to
our Tevatron $b$-tagging efficiency to reproduce the efficiencies
quoted in the experimental analyses.

The choice of cuts to distinguish signal from background is guided by
a few key features.  SM backgrounds exhibit $\met$ either because of
neutrino production or jet energy mismeasurement.  The first source is
suppressed by a lepton veto, which rejects processes involving $W \to
\ell \nu$.  The second source can be effectively suppressed by a
combination of minimum $\met$ cuts (which suppress multi-jet
backgrounds) and the requirement that $\met$ not be aligned with any
energetic jets.  The alignment cut is imposed in terms of a minimum
angle between the missing transverse energy and any of the selected
jets, $\Delta\phi_{\rm min}(\met, {\rm jets})$.  Since $\met$ can also
be mismeasured in events with significant transverse momentum in
low-energy jets and leptons, such events can be removed by cuts on the
quantity ${\cal A} \equiv (\met - \mht)/(\met + \mht)$, where $\mht
\equiv |\sum_{\rm jets} \vec p_T^j|$ is the magnitude of the vector
sum of the transverse energy of all jets.

Moreover, signal events result in the pair production of two massive
objects, whose decays to $b$-jets and invisible particles are roughly
uncorrelated.  These events are thus expected to have several objects
with large $p_T$, with most of the transverse momentum carried by the
two leading jets, which are not expected to be back-to-back.  The
$\met$ is also expected to be comparable to the $p_T$ of the other
objects.  For the Tevatron, the presence of objects with large $p_T$
is measured by the kinematic variable $H_T \equiv \sum_{\rm jets}
p_T^j$, while at the LHC it is measured by the kinematic variable
$M_{\rm eff} = \met +\sum_{j_1 ... j_4} p_T^j$.

At the Tevatron, the requirement that the leading jets not be
back-to-back is measured by the kinematic variable $\alpha_{j_1 j_2}$,
defined as the angle between the two leading jets in the transverse
plane.  At the LHC this requirement is imposed in terms of the
transverse sphericity, $S_T$.  If $\lambda_{1,2}$ are the eigenvalues
of the $2 \times 2$ sphericity tensor $S_{ij}=\sum_k p_{ki} p^{kj}$
for all selected jets, one defines $S_T \equiv 2\lambda_2 / (\lambda_1
+ \lambda_2)$.  QCD backgrounds are dominated by back-to-back jet
configurations, for which $S_T \sim 0$.

At the Tevatron, the requirement that most transverse momenta be
carried by the two leading jets is measured by the kinematic variable
$X_{jj} \equiv (p_T^{j_1}+p_T^{j_2})/H_T$.  At the LHC, the
requirement that the missing transverse energy be comparable to the
momenta of other objects is measured in terms of $f \equiv \met /
M_{\rm eff}$.

Note that we have chosen not to take into account next-to-leading
order $K$ factors in our analysis. We consider this conservative, in
the sense that next-to-leading calculations tend to increase signal
and background (both QCD and vector boson + jets) with a similar
factor ($\sim 1.2-1.5$), and therefore tend to increase the
significance of the result and improve the exclusion and discovery
regions \cite{Campbell:2002tg, Cacciari:2008zb}.

\subsection{Tevatron Cuts}

For Tevatron searches, we begin by imposing precuts that are similar
to the cuts required in Ref.~\cite{Abazov:2010wq} for sbottom
searches.  With these precuts, we require:

\begin{itemize}
\setlength{\itemsep}{1pt}\setlength{\parskip}{0pt}\setlength{\parsep}{0pt}
\item{}0 lepton with $|\eta_\ell| \leq 2.0$ and $p_T^{e,\mu} \geq
15~\gev$ and $p_T^{\tau} \geq 10~\gev$.
\item{}2 or 3 jets with $|\eta_j| \leq 2.5$ and $p_T^{j}\geq 20~\gev$.
\item{}$\alpha_{j_1j_2} \leq 165^\circ$.
\item{}$\met \geq 40~\gev$, $\met/\gev \geq 80 - 40 \times
\Delta\phi_{\rm min}(\met, {\rm jets})$.
\item{} At least two jets, including the leading jet, are tagged as
$b$-jets.
\item{} $\Delta\phi_{\rm min}(\met, {\rm jets}) \geq 0.6$ rad.
\item{} $-0.1 < {\cal A} < 0.2$.
\item{} $X_{jj} \geq 0.75$.
\end{itemize}

In addition, for each grid point in ($m_{B'}$, $m_X$) space, we
consider the following cuts and choose the combination that optimizes
the signal's significance:
\begin{itemize}
\setlength{\itemsep}{1pt}\setlength{\parskip}{0pt}\setlength{\parsep}{0pt}
\item{} $p_T^{j_1} \geq$ 50, 80, 100, 150 GeV.
\item{} $\met \geq$ 100, 150, 200, 250 GeV.
\item{} $X_{jj} \geq$ 0.9.
\item{} $H_T \geq$ 150, 220, 300 GeV.
\end{itemize}

Note that some of the final cut combinations are redundant (e.g., $H_T
< p_T^{j_1}+p_T^{j_2}$), resulting in a total of 160 naively
non-redundant combinations.

Comparing our cut efficiencies for sbottom pair production with those
listed in Ref.~\cite{Abazov:2010wq} for two signal benchmark points
$(m_{\tilde{b}_1}, m_{\chi^0_1})=(240~\gev, 0~\gev)$ and $(130~\gev,
85~\gev)$, we found reasonably good agreement, except for the
$b$-tagging efficiency for two jets, which was underestimated by about
a factor of 2 in PGS4.  We therefore increase our cut efficiencies by
a factor of two (for both the signal process and the $t\bar{t}$
background) to account for this underestimation caused by our using
PGS4 for detector simulation.

\subsection{LHC Cuts}

For the LHC, we adopt the following precuts based on the cuts designed
for inclusive SUSY searches for 0 lepton, 2$-$3 jets (including 1$-$2
$b$-jets) at the LHC~\cite{ATLAS7,ATLAS10,Aad:2009wy}:

\begin{itemize}
\setlength{\itemsep}{1pt}\setlength{\parskip}{0pt}\setlength{\parsep}{0pt}
\item{}0 lepton with $|\eta_{e,\mu ,\tau}| \leq 2.5$ and
$p_T^{e,\mu,\tau} \geq 20~\gev$.
\item{}2 or 3 jets with $|\eta_j| \leq 2.5$ and $p_T^{j_1}\geq 100$
GeV, $p_T^{j_{2,3}}\geq 40~\gev$, $p_T^{j{\rm (veto)}}= 30~\gev$.
\item{}$\met \geq 80~\gev$.
\item{} $f\equiv \met/M_{\rm eff}$, $f \geq$ 0.3 (0.25) for 2-jet (3-jet)
events.
\item{} $\Delta\phi_{\rm min}(\met, {\rm jets}) \geq 0.2$ rad for all
selected jets.
\item{} Transverse sphericity $S_T \geq$ 0.2.
\item{} At least one selected jet is tagged as a $b$-jet.
\end{itemize}

The cuts on the transverse momentum of jets are chosen to satisfy the
trigger requirements, as well as to reject a sufficient amount of QCD
jet background.

In addition, for each grid point in ($m_{B'}$, $m_X$) space, we
consider the following cuts and choose the combination that optimizes
the signal's significance:
\begin{itemize}
\setlength{\itemsep}{1pt}\setlength{\parskip}{0pt}\setlength{\parsep}{0pt}
\item{} $p_T^{j_1} \geq$ 150, 200, 250, 300 GeV.
\item{} $\met \geq$ 100, 150, 200, 250, 300 GeV.
\item{} $M_{\rm eff} \geq$ 250, 300, 400, 500, 600, 700 GeV.
 \end{itemize}

Note that some of these final cut combinations are redundant (e.g.,
when $M_{\text{eff}} < p_T^{j_1}+p_T^{j_2}+\met$), resulting in a
total of 104 non-redundant combinations.

\section{Results and Discussion}
\label{sec:results}

\begin{figure}[tb]
\begin{center}
\includegraphics*[width=0.49\columnwidth]{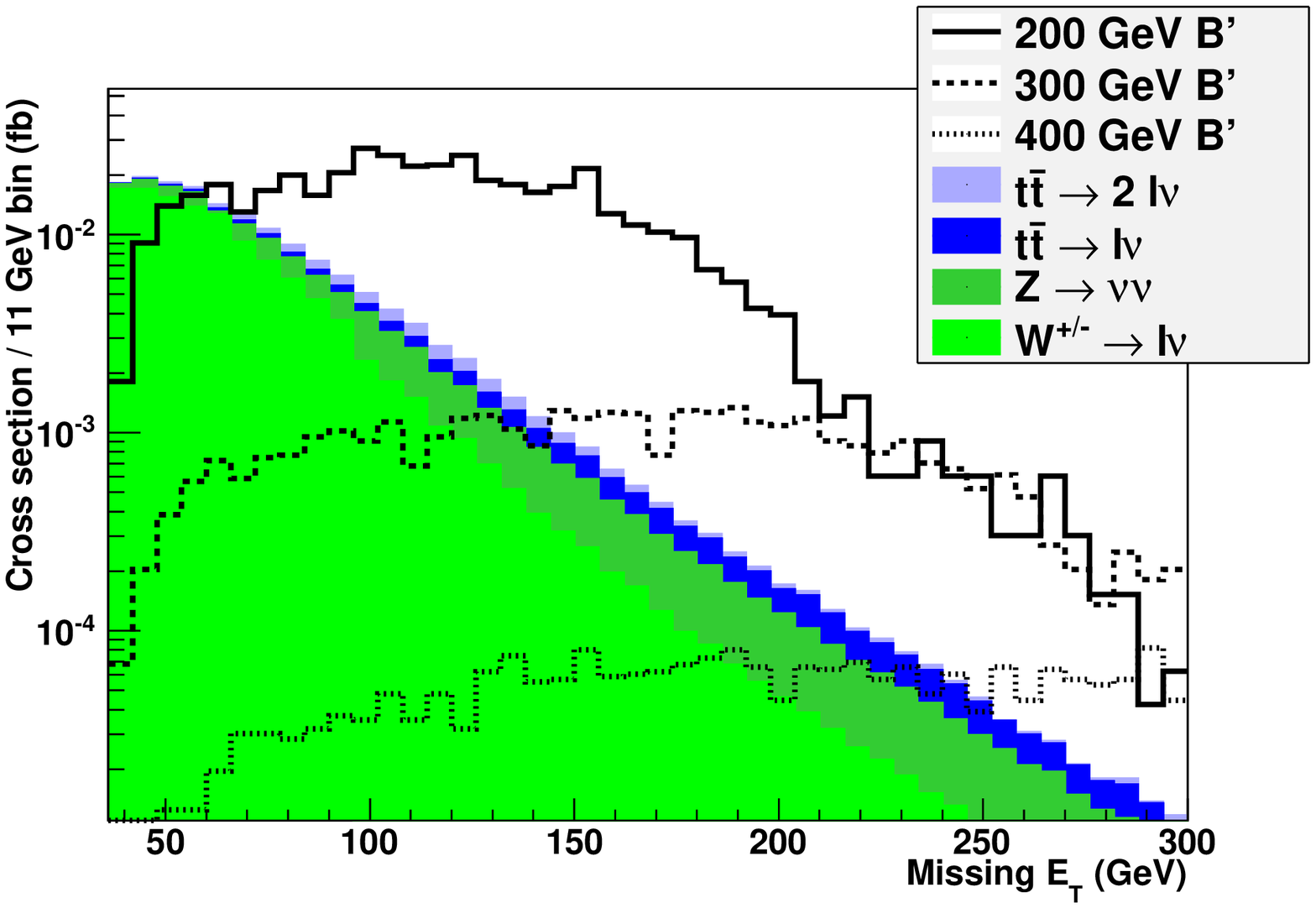}
\includegraphics*[width=0.49\columnwidth]{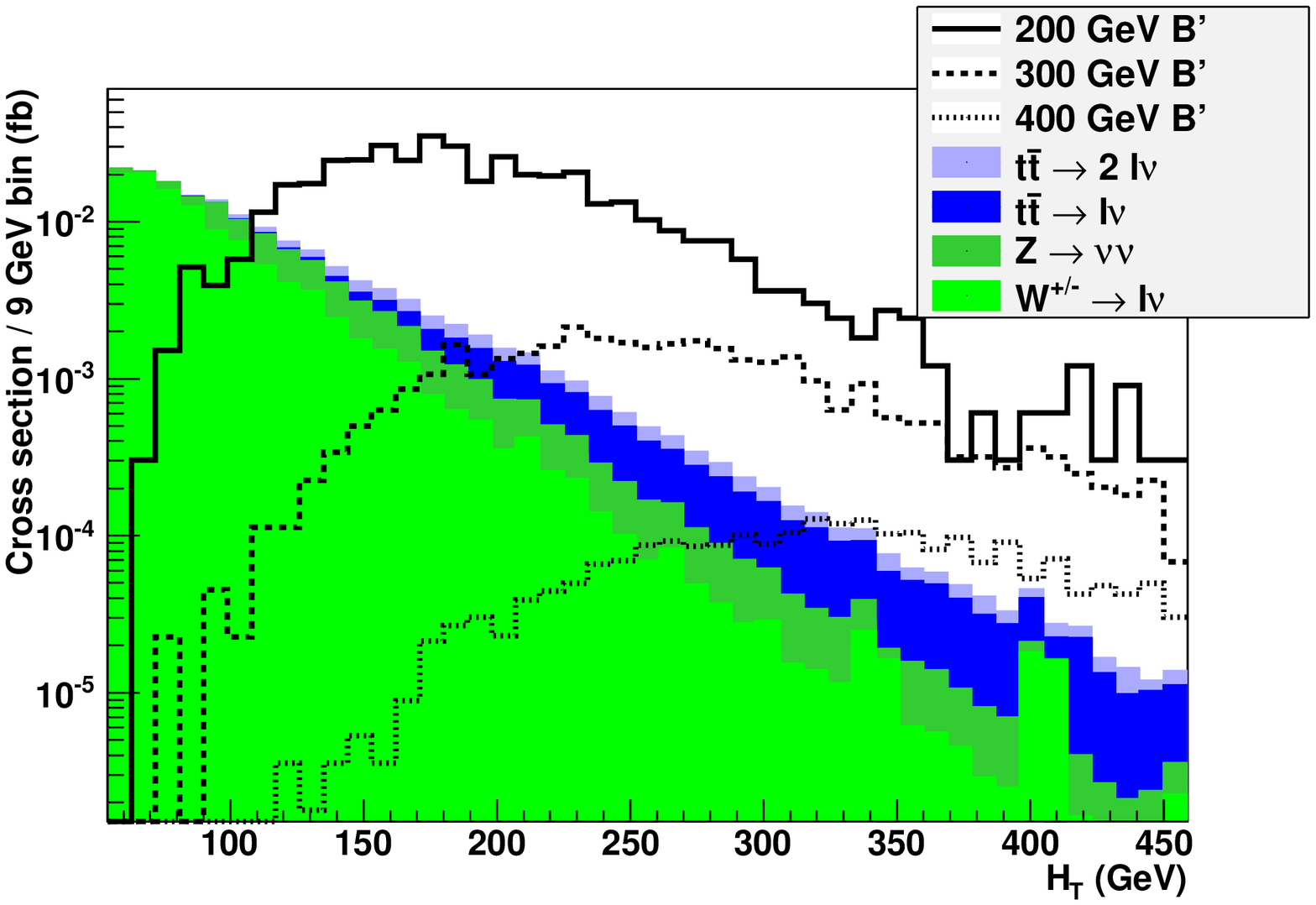}
\end{center}
\vspace*{-.25in}
\caption{\label{fig:distributions-tev} $\met$ and $H_T$ distributions
at the Tevatron for SM backgrounds and three signal benchmark points
$(m_{B'}, m_X)=(200,1),\ (300,1),\ (400,1)$ GeV, after
precuts.}
\label{fig:tevmetpt}
\end{figure}

\Figref{tevmetpt} shows the $\met$ and $H_T$ distributions for three
benchmark points $(m_{B'}, m_X)=(200,1),\ (300,1),\ (400,1)$ GeV as
well as the dominant SM backgrounds at the Tevatron after precuts.
$W(\ell\nu)jj$ becomes the dominant background after precuts, and
$Z(\nu\nu)jj$ and semileptonic $t\bar{t}$ background are relatively
large as well. While the differential cross section distributions for
the SM backgrounds drop quickly with increasing $\met$ and
$p_T^{j_1}$, the distributions for the signal typically extends to
much larger values of $\met$ and $H_T$, given the relatively large
mass splittings between $m_{B'}$ and $m_X$.  As a result, additional
cuts on $\met$ and $H_T$ (as well as $p_T^{j_1}$ and $X_{jj}$) can
effectively suppress the backgrounds while keeping most of the signal
intact, thereby optimizing the signal significance.

\begin{figure}[tb]
\begin{center}
\includegraphics*[width=0.49\columnwidth]{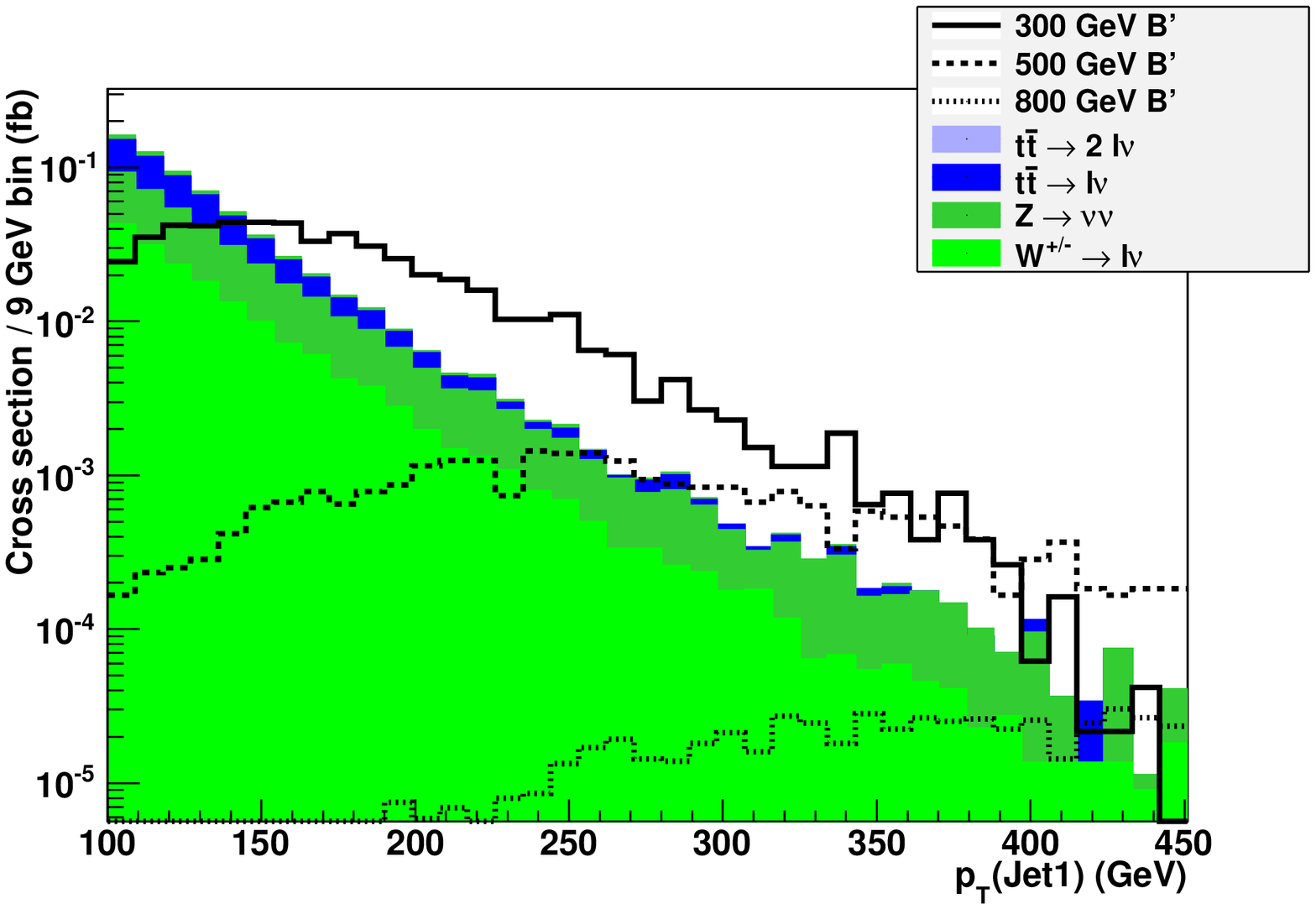}
\includegraphics*[width=0.49\columnwidth]{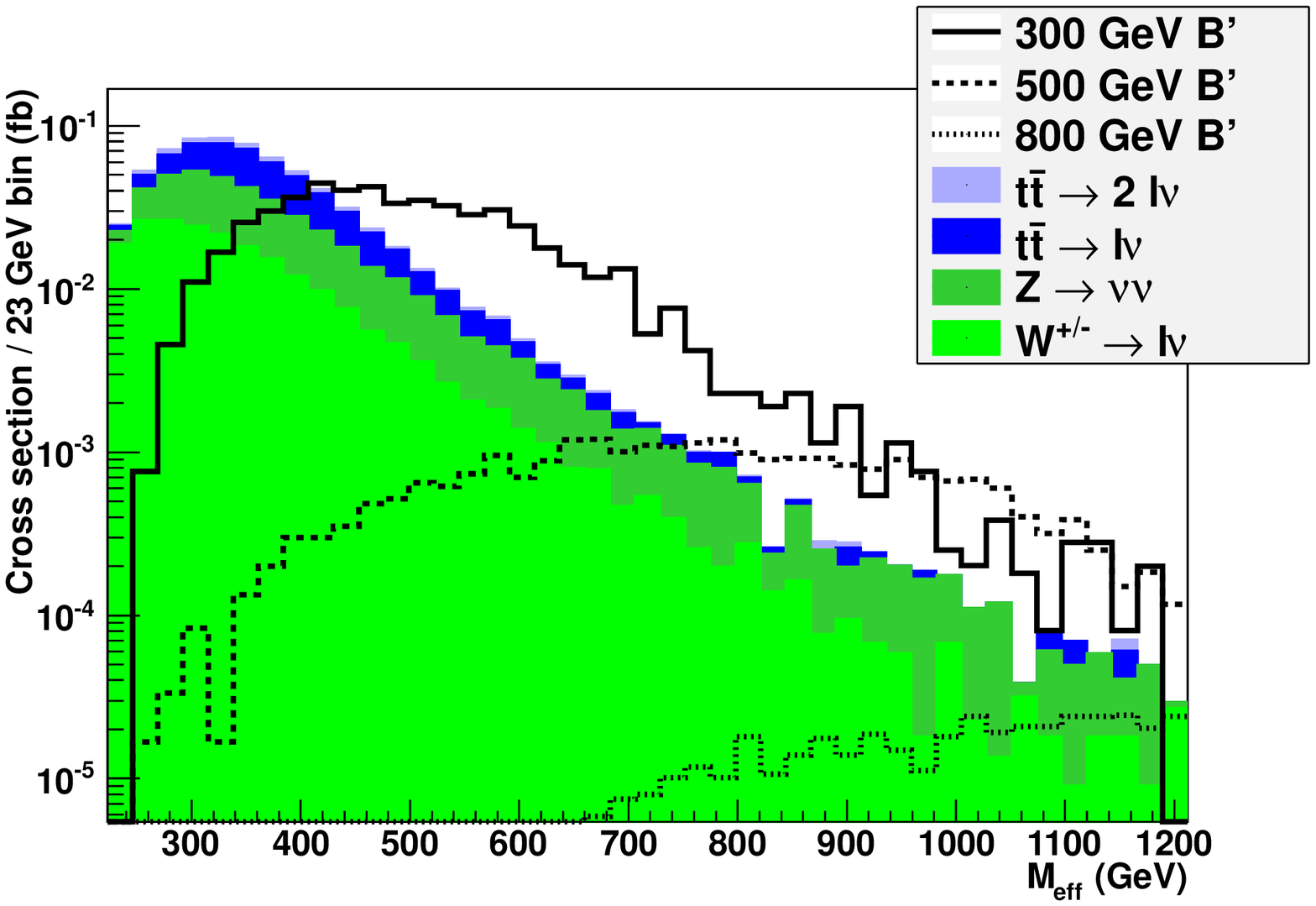}
\end{center}
\vspace*{-.25in}
\caption{\label{fig:distributions-lhc7} $p_T^{j_1}$ and $M_{\rm eff}$
distributions at 7 TeV LHC for SM backgrounds and three signal
benchmark points $(m_{B'}, m_X)=(300,1),\ (500,1),\ (800,1)$
GeV, after precuts. }
\label{fig:lhcmetpt}
\end{figure}

Similarly, $p_T^{j_1}$ and $M_{\rm eff}$ distributions at the 7 TeV
LHC after precuts are shown in \figref{lhcmetpt} for three benchmark
points $(m_{B'}, m_X)=(300,1),\ (500,1),\ (800,1)$ GeV, as well as the
dominant SM backgrounds.  The contributions from $Z(\nu\nu)jj$,
$W(\ell \nu)jj$ and $t\bar{t}$ are similar after precuts.  However,
for $t\bar{t}$, the differential cross section distributions drop
quickly with increasing $\met$, $p_T^{j_1}$ and $M_{\rm eff}$, since
the $t\bar t$ distributions are enhanced in the region below the $t$
mass (for $\met$ and $p_T^{j_1}$) and $2m_t$ (for $M_{\rm eff}$).
With additional cuts on $\met$, $p_T^{j_1}$ and $M_{\rm eff}$, the
$t\bar{t}$ background is almost negligible.  The $Z(\nu\nu)jj$
background, on the other hand, becomes dominant once additional cuts
are imposed.  For the signal benchmark point $(m_{B'}, m_X)=(300,1)$
GeV, the distributions drop quickly above the mass scale of the $B'$.
To optimize the cuts for such low $m_{B'}$, usually no additional
$\met$ or $p_T^{j_1}$ cuts are needed and the $M_{\rm eff}$ cut
becomes the most effective in selecting the signal. For larger masses,
the $p_T^{j_1}$ and $\met$ cuts become very effective in suppressing
the backgrounds. The cross sections for signal benchmark points and SM
backgrounds after various stages of cuts are presented in the
Appendix.

We now determine the discovery and exclusion reach for $B'$ at the
Tevatron and the 7 TeV LHC. For each parameter point
$(m_{B'},m_X)$, we use the optimum cut (after precuts) that
gives the best signal significance, with the additional requirements
that $S/B > 0.1$ and more than two signal events are observed. Given
the small number of signal and background events after cuts, we have
used Poisson statistics, rather than assuming Gaussian distributions,
for both signal and backgrounds.

\begin{figure}[tb]
\begin{center}
\includegraphics*[width=0.49\columnwidth]{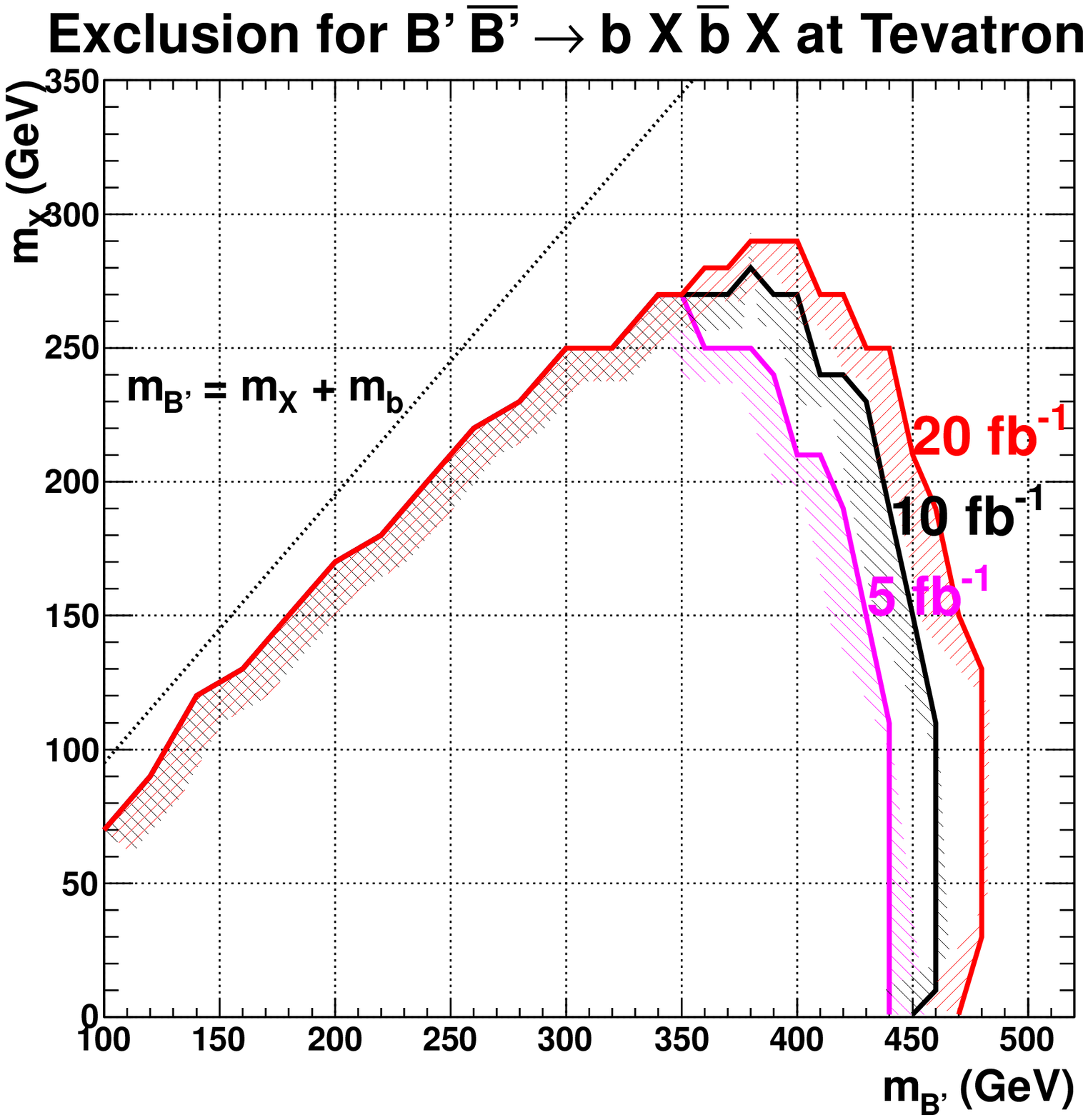}
\includegraphics*[width=0.49\columnwidth]{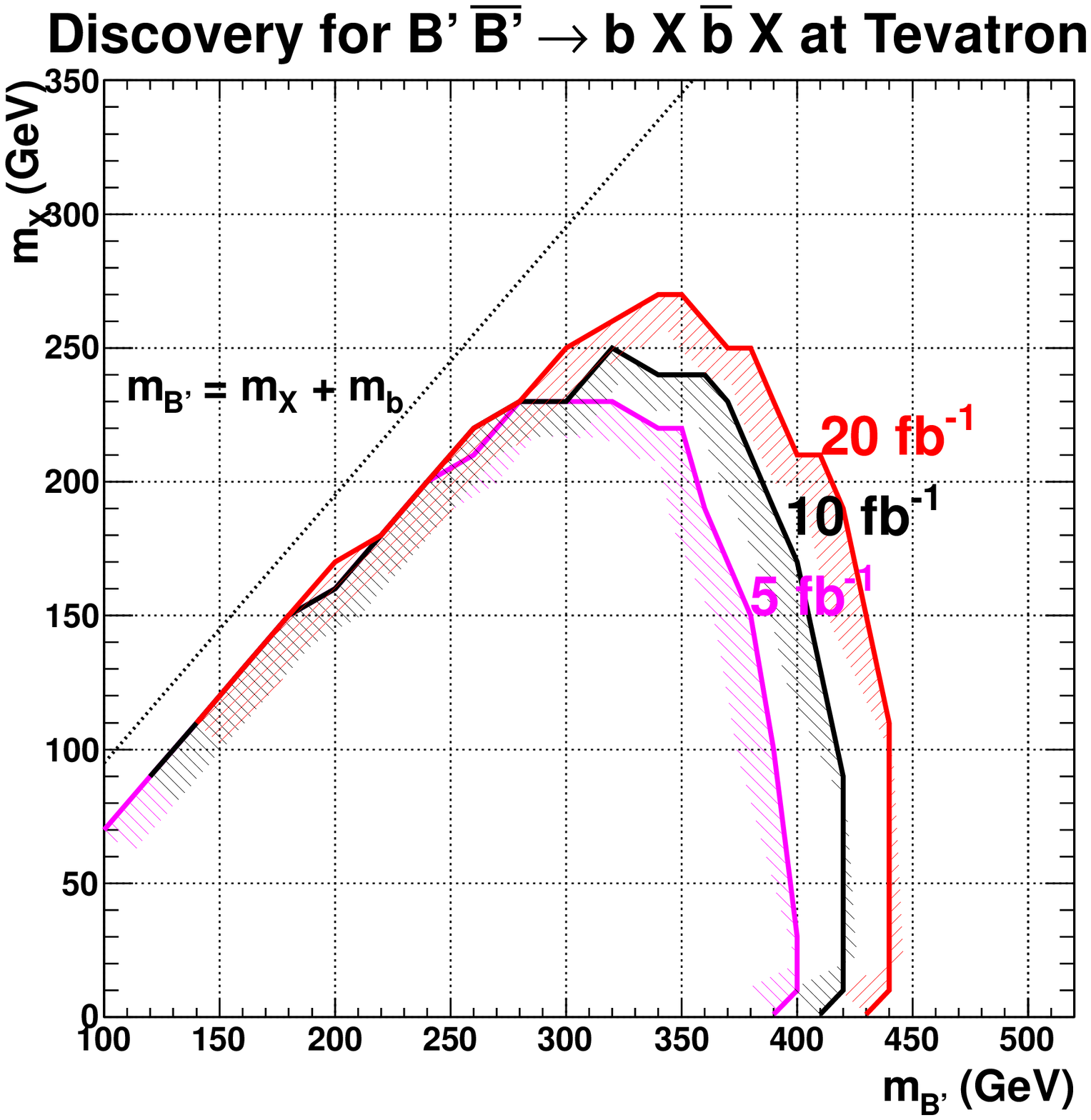}
\end{center}
\vspace*{-.25in}
\caption{95\% CL Tevatron exclusion (left plot) and 3 $\sigma$
discovery (right plot) reach in the $(m_{B'},m_X)$ plane for
integrated luminosities 5, 10, and $20~\ifb$. For each point in
parameter space, the cut with the best significance has been chosen.}
 \label{fig:tevreach}
\end{figure}

\Figref{tevreach} shows the 95\% CL Tevatron exclusion and
$3\sigma$ (Gaussian equivalent\footnote{By Gaussian equivalent, we
mean that we have converted the one-sided Poisson probability into the
equivalent $\sigma$ deviation in a two-sided Gaussian distribution,
which is more commonly used in the literature.}) discovery contours in
the $(m_{B'}, m_X)$ plane.  For relatively small values of $m_X$,
$m_{B'}$ could be excluded up to 440, 460 and 480 GeV, or could be
discovered at $3\sigma$ up to 400, 420 and 440 GeV for integrated
luminosities of 5, 10, and $20~\ifb$, respectively.  For small mass
splittings $m_{B'}-m_X$, the $b$-jets become soft and the amount of
transverse missing energy gets smaller.  It is more challenging to
select signals out of the SM backgrounds with such soft decay
products.  This explains the gap between the exclusion/discovery
contours and the dashed line, which corresponds to the threshold for
the on-shell decay of $B' \to b X$.  With 20 ${\rm fb}^{-1}$
integrated luminosity, masses $m_X$ as large as 290 GeV may be
excluded, and masses $m_X$ as large as 270 GeV may be discovered.

\begin{figure}[tb]
\begin{center}
\includegraphics*[width=0.49\columnwidth]{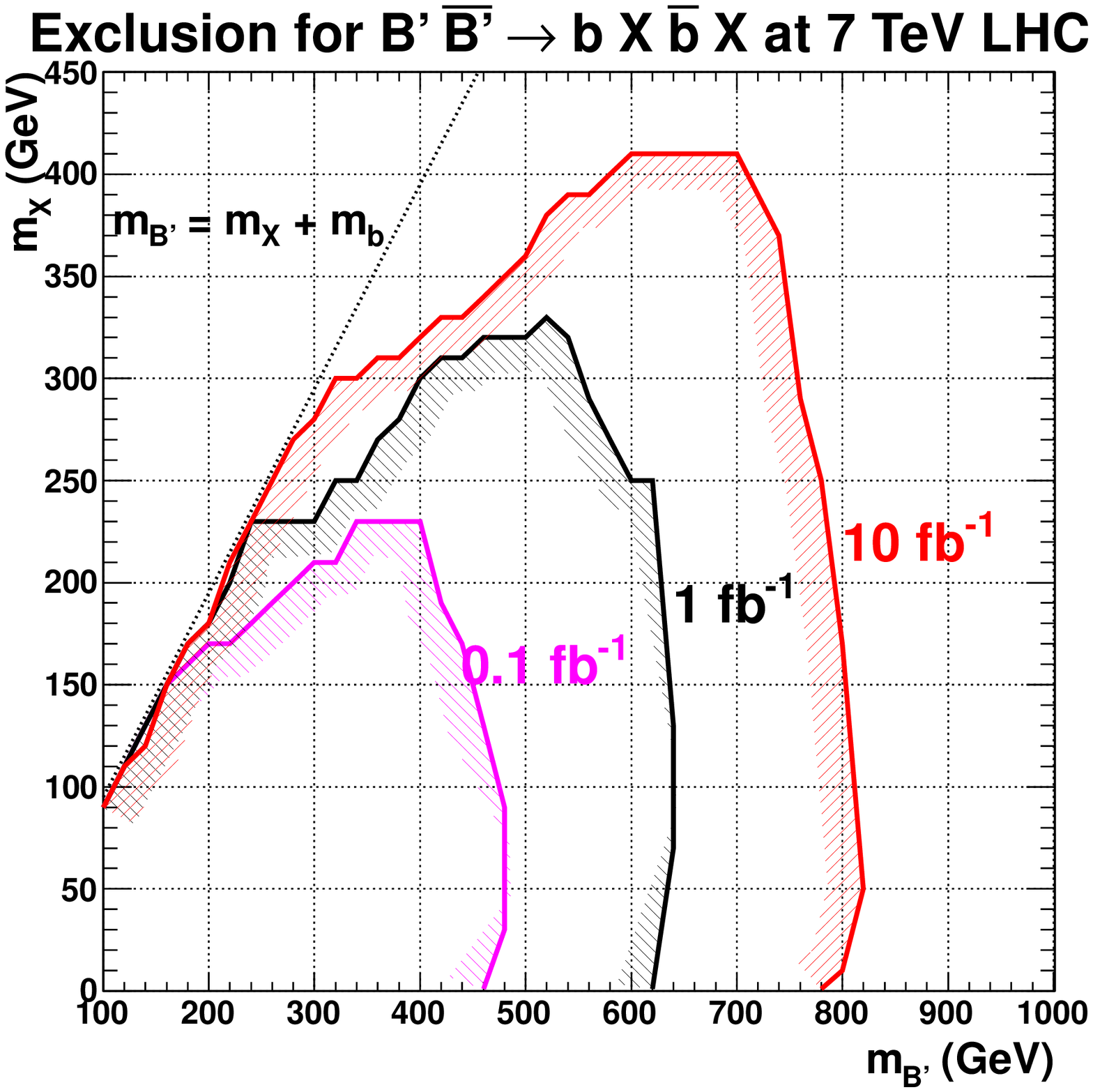}
\includegraphics*[width=0.49\columnwidth]{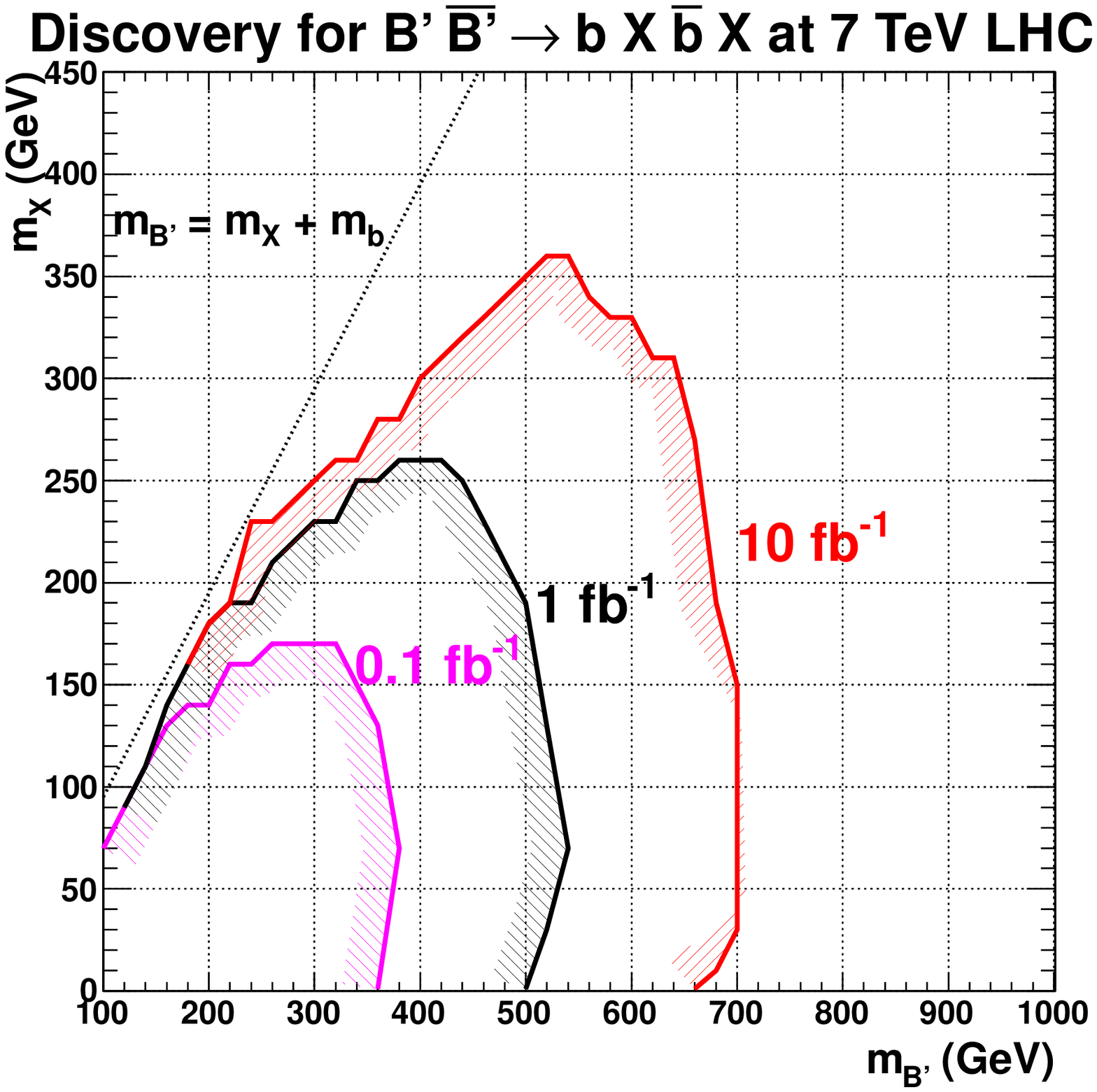}
\end{center}
\vspace*{-.25in}
\caption{95\% CL LHC 7 TeV exclusion (left plot) and 3 $\sigma$
discovery (right plot) reach in the $(m_{B'},m_X)$ plane for
integrated luminosities 0.1, 1, and $10~\ifb$. For each point in
parameter space, the cut with the best significance has been chosen. }
 \label{fig:LHC7reach}
\end{figure}

\Figref{LHC7reach} shows the 95\% CL exclusion and 3$\sigma$
(Gaussian equivalent) discovery contours for a 7 TeV early LHC run,
for integrated luminosities 0.1, 1, and $10~\ifb$.  With just
$0.1~\ifb$, the LHC exclusion reach for $m_{B'}$ of 480 GeV exceeds
the Tevatron exclusion reach with $20~\ifb$ luminosity.  With $1~\ifb$
and $m_X \sim 0$, all regions of $m_{B'}$ in the perturbative Yukawa
coupling region can be covered.  Exclusions of $m_{B'}$ up about 800
GeV could be achieved with $10~\ifb$ integrated luminosity.  Note that
at the LHC, we could tolerate a much smaller $m_{B'}-m_X$ almost up to
the on-shell decay threshold for small $m_{B'}$.  The 3$\sigma$
discovery reach for $m_{B'}$ is about 380, 540, 700 GeV for integrated
luminosities of 0.1, 1, and $10~\ifb$.  The reach in $m_X$ is greatly
enhanced at the LHC as well.  It could be excluded up to 330 and 410
GeV, or to be discovered up to 260 and 360 GeV with 1 and 10 $\ifb$
data.

Note that the exclusion curve for the Tevatron in \figref{tevreach}
fails to reach the $m_X=m_{B'}-m_b$ line for any values of $m_{B'}$,
because of the small missing energy in this region.  As evident in
\figref{LHC7reach}, however, this is not true at the LHC, where the
energy and cross section for low-mass $B'$s are so large that recoil
of the $B'\bar{B}'$ system against initial state radiation jets gives
events with sufficient $\met$ to pass the cuts, even for
$m_X=m_{B'}-m_b$. For further discussion, see
Ref.~\cite{Alwall:2008ve}.

\begin{figure}[tb]
\begin{center}
\includegraphics*[width=0.49\columnwidth]{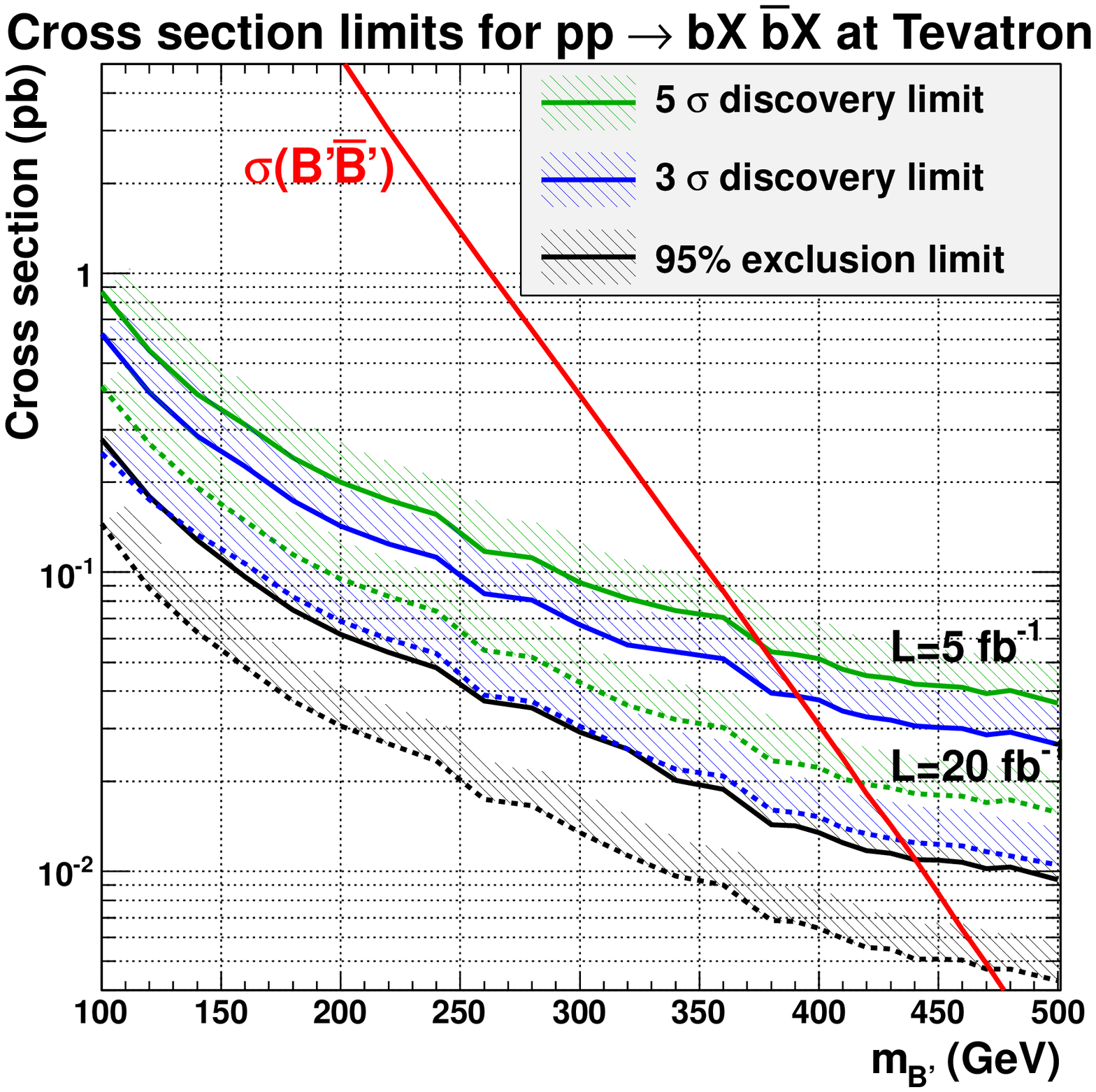}
\includegraphics*[width=0.49\columnwidth]{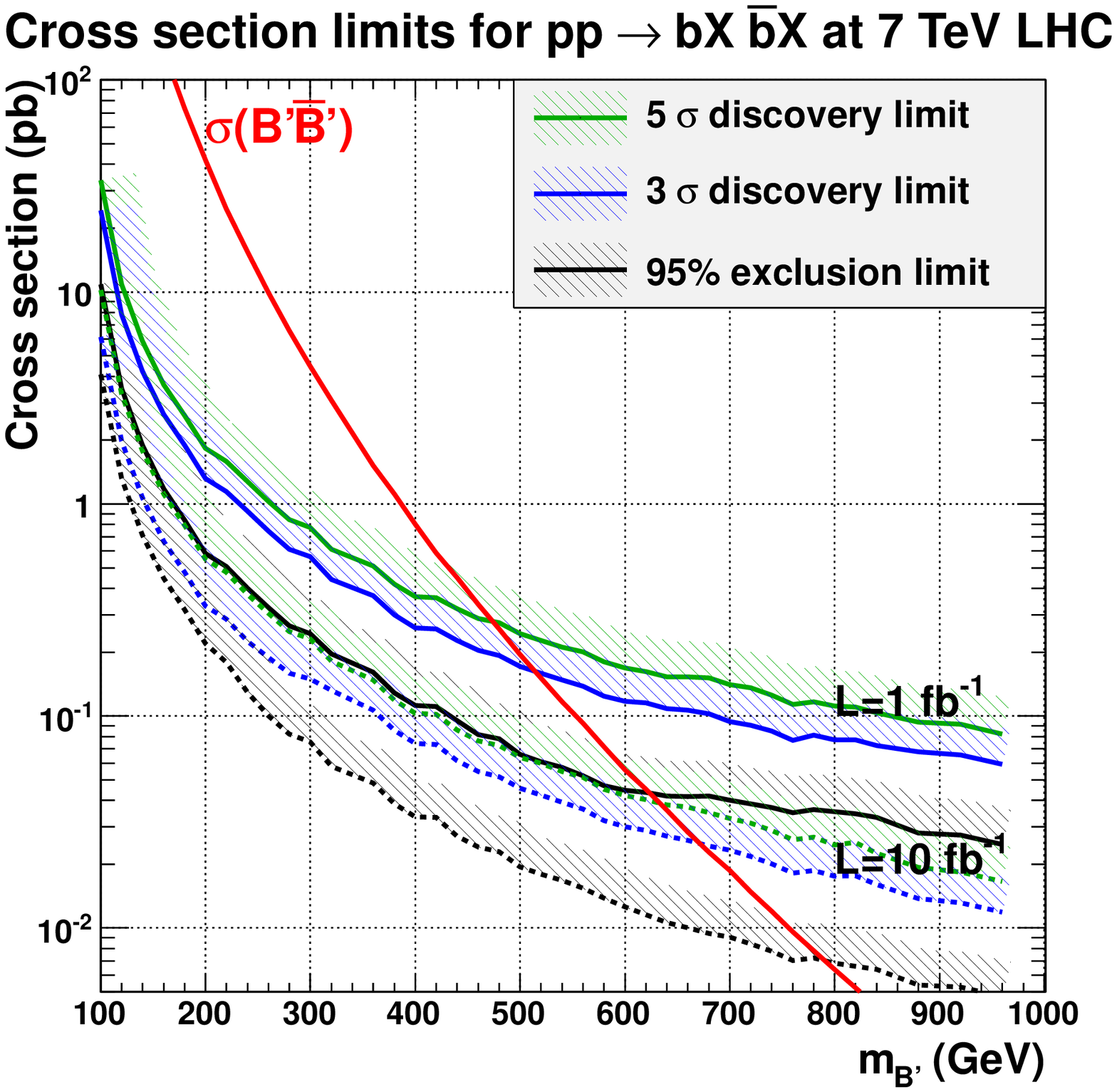}
\end{center}
\vspace*{-.25in}
\caption{Model-independent 95\% exclusion, 3$\sigma$ and 5$\sigma$
discovery cross section reaches for light $m_X$ ($m_X=1$ GeV) at the
Tevatron (left plot) and 7 TeV LHC (right plot).  Also shown are the
QCD pair production cross sections for $B'\bar{B}'$ (solid
red curves).  }
 \label{fig:cross_section_limits}
\end{figure}

To present our exclusion and discovery reaches in a more
model-independent way, in \figref{cross_section_limits} we show the
collider reaches of the $bX \bar{b}X$ production cross section [or
equivalently, $\sigma(B' \bar{B}')\times B(B' \to b X)^2$] as a
function of $m_{B'}$ for 95\% CL exclusion, 3$\sigma$ and 5$\sigma$
discovery for various luminosities at the Tevatron and the 7 TeV LHC,
with $m_X$ fixed to 1 GeV.  At the Tevatron with 20 $\ifb$, production
cross sections of $5-$200 fb could be excluded at 95\% CL for the mass
of $B'$ in the range of 100$-$500 GeV.  The limits get better for
higher masses due to the more energetic final state particles and
large $\met$ and $M_{\rm eff}$ in the signal process.  For $5\sigma$
discovery, the cross section reach is about 20$-$400 fb.  At the 7 TeV
LHC, with $1~\ifb$ luminosity we could reach an exclusion limit of
about 25 fb for $m_{B'}$ around 1 TeV.  A $5\sigma$ reach of 20 fb can
be achieved with $10~\ifb$ luminosity.

For the purpose of illustration, we also show the QCD pair production
cross sections of $B'\bar{B}'$ (solid curves) in
\figref{cross_section_limits}.  For $B'\bar{B}'$ pair production in
other new physics models, one can easily read out the collider reach
of $m_{B'}$ by comparing $\sigma(B' \bar{B}')\times B(B' \to b X)^2$
in those models with the cross section reach curves in
\figref{cross_section_limits}.

Note that we have taken $m_X=1$ GeV when presenting the cross section
reaches at colliders.  However, as evident from
\figsref{tevreach}{LHC7reach}, the reach in $m_{B'}$ is almost
independent of $m_X$ for small and moderate values of $m_X$, unless
the mass splitting of $m_{B'}-m_X$ becomes small.

\section{Conclusions}
\label{sec:conclusions}

In this study, we have considered the possibility of pair production
of new charge $-\frac{1}{3}$ quarks $B'$ that decay directly to
$b$-quarks and long-lived neutral particles $X$.  The resulting signal
is $B' \bar{B}' \to b \bar{b} \met$, which is common to many new
physics theories, as discussed in \secref{models}.  Since the $b
\bar{b} \met$ signal is common to many models, we have also presented
detection prospects in terms of the pair production cross-section
$\sigma(b \bar{b} XX )$, for varying $m_{B'}$ and various integrated
luminosities.  This analysis thus accommodates many models in which
there are new contributions to $B' \bar{B'}$ production, as well as
models where $B'$ is not spin-${1\over 2}$, as in the case of bottom
squarks.\footnote{If $B'$ is not spin-${1\over 2}$, then the angular
distributions of the outgoing jets will change, altering cut
efficiencies.  In this case, the comparison is only approximate.}

We have estimated the sensitivity of the Tevatron and LHC to new
physics resulting in this signal.  Currently published bounds have
been summarized in detail in \secref{limits}; very roughly, however,
and translating bounds on related processes to the case of $B'
\bar{B}'$ production, the current limits are $m_{B'} \agt 370~\gev$.
{}From our analysis, we expect that these results may be improved to
$m_{B'} \agt 440~\gev$ for small $m_X$, given our optimized cuts and
$5~\ifb$ of data, which is currently available at the Tevatron.  We
also find that additional Tevatron data will marginally improve this
bound: for $20~\ifb$ of data, models with $m_{B'} \leq 480~\gev$ ($m_X
\alt 150~\gev$) can be excluded at 95\% CL.

These results may be further improved at the LHC with an analysis of
$1~\ifb$ of data, which has already been accumulated.  With $1~\ifb$
of integrated luminosity, the LHC physics run can probe any models
with $m_{B'} \alt 620~\gev$, provided $m_X \alt 250~\gev$.  Models
with $B'$ quarks that get mass from electroweak symmetry breaking are
bounded by the requirement of perturbative Yukawa couplings to have
masses below this mass.  Early LHC data may therefore probe the full
range of possible quark masses in these models.  In particular, the
early LHC will probe all WIMPless models that could explain the data
of DAMA and CoGeNT (assuming dominant coupling to 3rd generation SM
quarks).  Even null results from the search discussed here will
therefore be of significant interest.

These same detection prospects are, of course, also applicable to
little Higgs models, or asymmetric dark matter models arising from
hidden sector baryogenesis.  For these models, however, theoretical
considerations provide no expected upper bound on $m_{B'}$.  The mass
reach which the LHC can achieve with greater luminosity is thus of
interest.  With $10~\ifb$ of data, the LHC can probe asymmetric dark
matter models and other similar frameworks at the $95\%$ CL for
$m_{B'} \alt 800~\gev$, provided $m_X \alt 200~\gev$.

UED models have perhaps the least constrained theoretical motivation,
since the $B'$ mass is not bounded by Yukawa coupling perturbativity
and relatively small $B'-X$ mass splittings are perfectly plausible.
With $10~\ifb$ the LHC can probe models with $m_X$ as large as
$410~\gev$ at the $95\%$ CL.  This maximum reach is obtained for
$m_{B'}$ in the $600-700~\gev$ range.

All of these detection prospects can be easily translated into mass
reaches for bottom squarks decaying directly to $b \chi^0_1$.  With
$1~\ifb$, the LHC can probe models with $m_{\tilde b} \leq 400~\gev$
at $95\%$ CL (provided $m_{\chi_1^0} \alt 150~\gev$).  With $10~\ifb$,
the LHC reach increases to $m_{\tilde b} \leq 520~\gev$.

It is worthwhile to compare the mass reach of this $B'$ search to that
of the $T'$ search examined in Ref.~\cite{Alwall:2010jc}.  There it
was found that, with $1~\ifb$ of data, the LHC could probe low-mass
dark matter models at $3\sigma$ for $m_{T'} \leq 490~\gev$.  The $B'$
search described here has similar reach for the same luminosity and
required signal significance.  However, the $T'$ detection prospects
were seen to drop rapidly with increasing $m_X$, with no sensitivity
at all expected for $m_X \geq 180~\gev$.  In contrast, the detection
prospects for this $B'$ search are almost unchanged for $m_X \alt
200~\gev$ (assuming $3\sigma$ significance).  This difference is
attributable to the large mass of the top quark; for relatively heavy
$X$, there is very little phase space left for the $T' \to tX$ decay.
Although the $t \bar{t} \met$ signals provides many more handles, in
the end, the naive expectation holds true: the reaches in $m_{B'}$ and
$m_{T'}$ are roughly similar, and for a fixed new quark mass, the dark
matter mass reach of the $B'$ search exceeds that of the $T'$ search
by roughly $m_t - m_b \approx 170~\gev$.

The analysis presented here determines the prospects for detecting an
excess in events with $b$-jets and missing $E_T$.  It is more
difficult to determine if the excess arises from the pair production
of $B'$, decaying via $B' \to bX$.  To determine the masses $m_{B'}$
and $m_X$ would be harder still.  It would be interesting to determine
the prospects for the LHC to make these measurements.

\section{Acknowledgments}

We are grateful to T.~Tait for useful discussions.  JA is supported by
Fermi Research Alliance, LLC under Contract No.~DE-AC02-07CH11359 with
the United States Department of Energy.  The work of JLF and SS was
supported in part by the National Science Foundation under
Grants~PHY-0653656 and PHY-0970173.  The work of JK was supported in
part by the Department of Energy under Grant~DE-FG02-04ER41291. The
work of SS was supported in part by the Department of Energy under
Grant~DE-FG02-04ER-41298.

\appendix

\section*{Appendix: Impact of Cuts on Signal and Backgrounds}
\label{sec:app}

In this Appendix, we present tables listing the cross sections after
cuts for the $B' \bar{B}'$ signal and the main SM backgrounds (Tables
\ref{table:TeVcuts}-\ref{table:LHC7cuts}).  In the upper section of
each table, each line gives the cross section after including all cuts
above.  In the lower section, each line gives the cross section after
including the cut on that line, and all precuts.  For the signal,
three examples with $m_X = 1~\gev$ and $m_{B'} = 200$, 300, and 400
(300, 500 and 800) GeV were chosen for the Tevatron (7 TeV LHC).  The
$W$ and $Z$ cross sections in parentheses were simulated with a cut on
$\met > 20\ (60)~\gev$ for the Tevatron (LHC) and at least 2 jets in
the parton-level generation.

\begin{table}[htbp]
\caption{Signal and background cross sections in pb after cuts for
signal and dominant backgrounds at the Tevatron.  The signal examples
are for $m_X = 1~\gev$ and $m_{B'} = 200$, 300, and 400 GeV as
indicated.  The $W$ cross sections in parentheses were simulated with
a cut on $\met > 20~\gev$ and at least 2 jets in the parton-level
generation. From the 160 independent combinations of final cuts used
for the cut optimization, the three cuts that optimize the
significance for these three mass points are displayed in the table.
Momenta and masses are in GeV.}  {\small
\begin{tabular}{|c|c|c|c|c|c|c|}
\hline
&$B'$ (200) & $B'$ (300) & $B'$ (400) & $W^\pm+ $jets & $Z\to\nu\nu+$jets & $t\bar t+$jets \\
\hline
No cut  & 2.62 & 0.195 & 0.0154 & (632.45) & (21.103) & 5.628\\
0 leptons  & 2.24 & 0.169 & 0.0134 & (229.22) & (16.516) & 2.365\\
$2 \le $ jets $\le 3$  & 1.89 & 0.143 & 0.0109 & (33.80) & (7.962) & 0.456\\
$\alpha_{j_1j_2} < 165^\circ$  & 1.66 & 0.125 & 0.0097 & (29.35) & (7.171) & 0.362\\
$\met > 40$  & 1.55 & 0.122 & 0.0096 & 17.05 & 5.221 & 0.235\\
$\met > 80 - 40\times \Delta \phi_{min}^{(\met, {\rm jets})}$  & 1.52 & 0.121 & 0.0095 & 16.45 & 5.042 & 0.207\\
$\Delta\phi(\met, {\rm jets}) > 0.6 $  & 1.43 & 0.112 & 0.0086 & 16.29 & 4.957 & 0.188\\
${\cal A} \equiv \frac{\met-\mht}{\met+\mht}$ cut  & 1.42 & 0.111 & 0.0086 & 15.89 & 4.869 & 0.179\\
$X_{jj} \equiv (p_T^{j_1}+p_T^{j_2})/H_T > 0.75$  & 1.25 & 0.102 & 0.0081 & 13.48 & 4.197 & 0.079\\
$p_T^{j_1} > 20$   & 1.25 & 0.102 & 0.0081 & 13.48 & 4.197 & 0.079\\
$H_T > 60$   & 1.25 & 0.102 & 0.0081 & 10.56 & 3.680 & 0.078\\
$\ge 2$ $b$-jets, $b$-jet hardest jet & 0.43 & 0.035 & 0.0026 & 0.11 & 0.037 & 0.018\\
\hline
All precuts & 0.43 & 0.035 & 0.0026 & 0.11 & 0.037 & 0.018\\
$X_{jj} > 0.9$, $\met > 40$,  $H_T > 300$  & 0.018 &  &  & $8.59\cdot10^{-5}$ & $1.11\cdot 10^{-4}$ & $1.68\cdot10^{-4}$\\
$X_{jj} > 0.9$, $\met > 150$,  $H_T > 300$  &  & 0.0043 &  & $4.56\cdot10^{-5}$ & $7.40\cdot10^{-5}$ & $4.11\cdot10^{-5}$\\
$X_{jj} > 0.9$, $\met > 250$,  $H_T > 300$  &  &  & $5.30\cdot10^{-4}$ & $1.95\cdot10^{-5}$ & $4.02\cdot10^{-5}$ & $1.39\cdot10^{-5}$\\
\hline
\end{tabular}
}\label{table:TeVcuts}
\end{table}

\begin{table}[htbp]
\caption{Signal and background cross sections in pb after cuts for
signal and dominant backgrounds at 7 TeV LHC.  The signal examples are
for $m_X = 1~\gev$ and $m_{B'} = 300$, 500, and 800 GeV as indicated.
The $W$ cross sections in parentheses were simulated with a cut on
$\met > 60~\gev$ and at least 2 jets in the parton-level
generation. From the 104 independent combinations of final cuts used
for the cut optimization, the three cuts that optimize the
significance for these three mass points are displayed in the table.
Momenta are in GeV units.  } {\small
\begin{tabular}{|c|c|c|c|c|c|c|}
\hline
Cut  & $B'$ (300) & $B'$ (500) & $B'$ (800) & $W^\pm+$ jets
& $Z\to\nu\nu+$ jets & $t\bar t$ \\
\hline
No cuts  & 4.47 & 0.195 & $6.39\cdot10^{-3}$ & (194.13) & (49.19) & 94.96 \\
0 leptons  & 4.05 & 0.179 & $5.83\cdot10^{-3}$ & (104.32) & (43.69) & 50.07 \\
$\ge 2$ jets, $p_T^{j_1} > 100$, & & & & & &   \\
veto 4th  jet at 30 GeV & 2.64 & 0.122 & $3.79\cdot10^{-3}$ & (18.50) & (12.33) & 6.99 \\
$\met > 80$  & 2.20 & 0.113 & $3.67\cdot10^{-3}$ & 13.55 & 9.77 & 1.66 \\
$f\equiv\met /M_{\rm eff} >$ 0.3 (2-jets)  & 1.75 & 0.090 & $3.00\cdot10^{-3}$ & 9.93 & 7.69 & 1.29 \\
$f\equiv\met /M_{\rm eff} >$ 0.25 (3-jets)  & 1.55 & 0.080 & $2.66\cdot10^{-3}$ & 9.30 & 7.37 & 0.94 \\
$\Delta\phi(\met, {\rm jets})>$ 0.2  & 1.50 & 0.077 & $2.49\cdot10^{-3}$ & 8.82 & 7.04 & 0.89 \\
$S_T > 0.2$  & 0.91 & 0.045 & $1.42\cdot10^{-3}$ & 3.79 & 3.21 & 0.52 \\
$\ge 1$ $b$-jet & 0.53 & 0.026 & $8.51\cdot10^{-4}$ & 0.21 & 0.22 & 0.26 \\
\hline
All precuts & 0.53 & 0.026 & $8.51\cdot10^{-4}$ & 0.21 & 0.22 & 0.26 \\
$p_T^{j_1} > 100$, $\met > 80$, $M_{\rm eff} > 600$  & 0.101 &  &  & $6.6\cdot10^{-3}$ & 0.012 & $4.0\cdot10^{-3}$ \\
$p_T^{j_1}> 250$, $\met > 300$, $M_{\rm eff} > 700$ &  & 0.010 &  & $1.0\cdot10^{-3}$ & $2.4\cdot10^{-3}$ & $1.5\cdot10^{-4}$ \\
$p_T^{j_1} > 400$, $\met > 80$, $M_{\rm eff} > 400$ &  &  & $4.7\cdot10^{-4}$ & $1.6\cdot10^{-4}$ & $3.7\cdot10^{-4}$ & $5.08\cdot10^{-5}$ \\
\hline
\end{tabular}
}\label{table:LHC7cuts}
\end{table}



\end{document}